# DeepFocus: Fast focus and astigmatism correction for electron microscopy

P.J. Schubert[1], R. Saxena[1], J. Kornfeld[1]

[1]Max Planck Institute for Biological Intelligence, Martinsried, 82152, Germany

## Abstract

High-throughput 2D and 3D scanning electron microscopy, which relies on automation and dependable control algorithms, requires high image quality with minimal human intervention. Classical focus and astigmatism correction algorithms attempt to explicitly model image formation and subsequently aberration correction. Such models often require parameter adjustments by experts when deployed to new microscopes, challenging samples, or imaging conditions to prevent unstable convergence, making them hard to use in practice or unreliable. Here, we introduce DeepFocus, a purely data-driven method for aberration correction in scanning electron microscopy. DeepFocus works under very low signal-to-noise ratio conditions, reduces processing times by more than an order of magnitude compared to the state-of-the-art method, rapidly converges within a large aberration range, and is easily recalibrated to different microscopes or challenging samples.

## Main

### Introduction

The high resolution of electron microscopy (EM), and the ability to image every sample detail, for tissue with the help of dense heavy-metal staining, remain unrivaled[1,2]. Massive improvements in automation allows the acquisition of 3D images of biological samples with nanometer resolution spanning millimeters[3,4]. While EM connectomics, the complete mapping of neuronal tissue, has been one of the key applications, automated 3D EM also enabled studies ranging from the analysis of cellular SARS-CoV-2 replication[5] to fuel cell research[6], demonstrating its wide applicability.

A key component of automated (3D) EM is to maintain high-quality images over the entire acquisition process, often involving millions of individual 2D images and 24/7 operations. This makes manual microscope parameter adjustments largely impossible. Automatic defocus and astigmatism correction algorithms remain a challenge despite their necessity, especially in high-throughput electron microscopy. This can be explained by sample diversity, the tight constraints on algorithm execution time, aberration correction convergence speeds, and low-electron dose budgets to avoid artefacts.

Existing solutions[7–10] in the area of scanning electron microscopy (SEM) are usually based on explicit physical models of the electron beam and its interaction with the sample (Fig. 1a). Measurements (images) with known perturbations are taken, followed by focus and stigmation parameter inference to estimate the wavefront aberrations. These physically grounded approaches, and those employing classical image sharpness scores[11–13], often struggle with generalization. In other words, they frequently fail to perform well on novel samples without expert parameter tuning. This tuning may be infeasible for users, particularly when the algorithm is integrated into the microscope control software.

A recent study introduced a complex approach that employs two artificial neural networks to evaluate the quality of SEM images and subsequently estimate working distance corrections using an updated state vector and a database comprising tens of thousands of manually labeled images[14]. Reinforcement learning was applied to the problem of electron beam alignment[15] and deep learning models were successfully used for focus correction in light microscopy[16,17]. Motivated by these developments and the outstanding performance of convolutional neural networks in general image processing tasks, we devised a new deep learning-based focusing and stigmatization correction method for scanning electron microscopy. Our algorithm features near-instant inference time, rapid convergence, functionality with low-electron dose noisy images, and a user-friendly process for recalibrating it to new machines and samples without the need for expert knowledge, ensuring convergence in all application scenarios.

## Results

The image of a flat specimen in a scanning electron microscope is optimally captured when the size of the spot of the electron beam is smaller than the sampling distance. Commonly, three parameters, working distance, on-axis stigmator and diagonal stigmator, henceforth referred to as stig x and y, can be adjusted by SEM operators to directly control the spot shape and bring it below the pixel size at the beam-specimen interaction point (Fig. 1a), consequently leading to sharp image formation (Fig. 1b).

The DeepFocus algorithm takes as input two SEM images with a known working distance perturbation $\sigma_{wd}$, around the current microscope working distance and stigmator settings $F = [f_{wd}\ f_{stig\ x}\ f_{stig\ y}]$, a technique known as phase diversity[18]. Multiple subregions (patches) are cropped from the two perturbed input images (Fig. 1c), and processed independently by a convolutional neural network. This network is trained to infer the $\Delta F$ that leads to a sharp image when added to $F$. The resulting multiple $\Delta F$ estimates, one for each input patch pair, are reduced by a mean operation, which serves as final output for a single iteration.

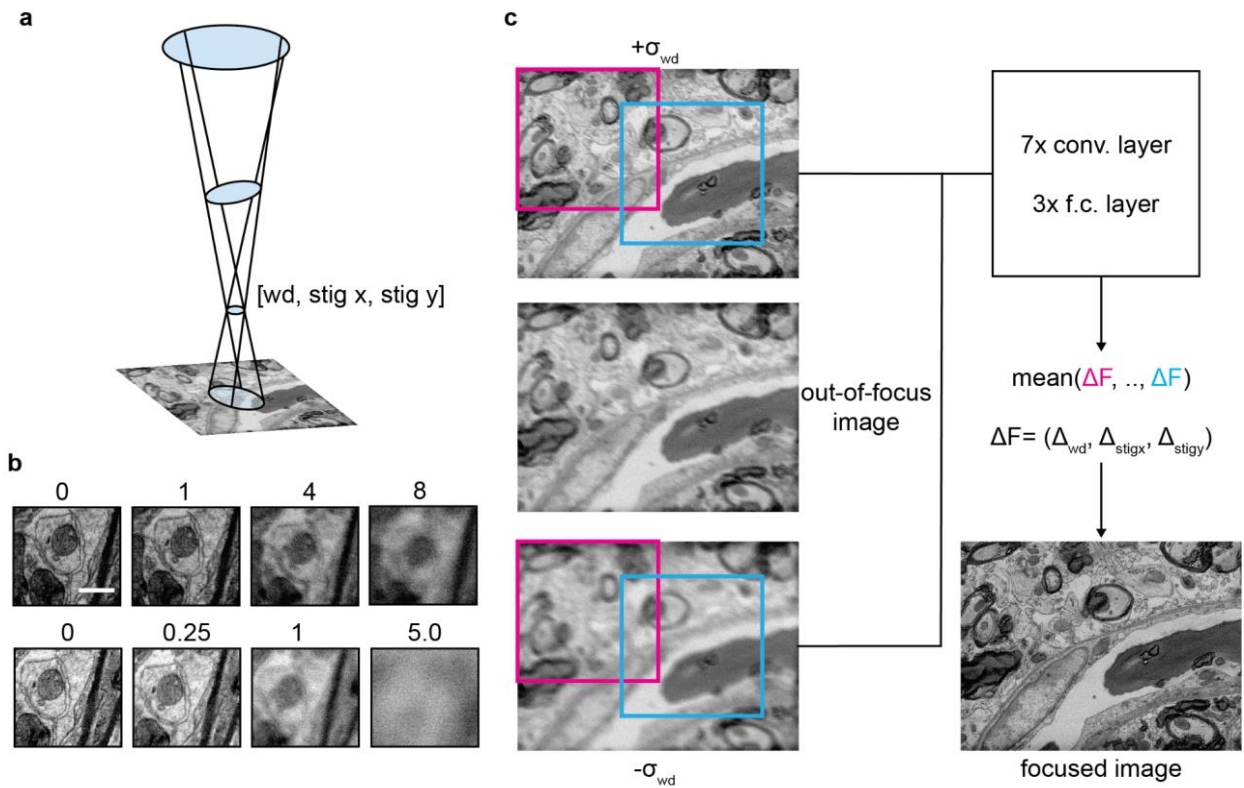

**Fig. 1** SEM beam formation and DeepFocus algorithm. **a** Schematic of the electron beam and the parameters that are controlled by DeepFocus. **b** Defocus - and astigmatism series that shows the influence of mild working distance (top row: 0 to 8 μm) and stigmator changes (bottom row: 0 to 5 a.u.) on image quality for a Zeiss Merlin SEM with 800 ns dwell time and 10 nm pixel size. Scale bar is 500 nm. **c** The out-of-focus image

(example resolution 1024 x 768, 10 nm pixel size) is perturbed (symmetric perturbation σ$_{wd}$ = ±5 μm) and N randomly located patch-pairs (shared offset within image pair, e.g. blue and pink squares with shape 512 x 512) of fixed shape are cropped and processed by DeepFocus (f.c.: fully connected, conv.: convolutional). The mean of N independent predictions is used to calculate a correction term ΔF for each focus parameter (wd: working distance, stig x: stigmator x; stig y: stigmator y). All SEM images have 10 nm pixel size. Scale bar in b is 500 nm.

To assess the model, we trained the network for about 44 hours on a single GPU on a set of 32 sample locations with different aberration parameters (in total n=320 input image pairs; Supplementary Fig. 1). We subsequently tested the model on location-aberration pairs that were not part of the training set (Methods). DeepFocus rapidly converges toward the target $\Delta\tilde{F}$ values within three iterations (Fig. 2a,b), even for low signal-to-noise (SNR) ratio image pairs (Fig. 2c,d) and small input patches (Supplementary Fig. 2a). We also examined the impact of input alignment, a strict requirement for example for the algorithm by Binding and Denk[7], and found that the model performs well also in the extreme case that the patches in an input pair did not share the same offset, but were chosen randomly (Supplementary Fig. 2b).

The average estimated correction $\Delta F = [\Delta f_{wd}\ \Delta f_{stig\,x}\ \Delta f_{stig\,y}]$ after a single iteration was assessed at nine distinct locations (evenly spaced grid with an edge length of 100 μm) for an expanded range of initial defocus (working distance perturbation in μm of ±20, ±10, ±5, ±2, ±1) to evaluate the model's learned transformation's goodness of fit. The relationship between the target correction for the working distance $\Delta\tilde{f}_{wd}$ (the negative introduced defocus) and model output $\Delta f_{wd}$ should ideally be linear (Fig. 2e), specifically, it should follow $\Delta f_{wd} = c_1 \cdot \Delta\tilde{f}_{wd} + c_2$ with $c_1 = 1$ and $c_2 = 0$. Using ordinary least squares (OLS; from the statsmodels Python package) to fit a line resulted in c1 = 0.9093 ± 0.006 and c2 = 0.3436 ± 0.061 (±1σ interval), signifying a slight yet significant deviation from the identity function.

Nevertheless, the model effectively learned to deduce the correction direction. The remaining mean absolute difference of the working distance $|\delta_{wd}| = |\Delta f_{wd} - \Delta\tilde{f}_{wd}|$ was closer to the target value $\Delta\tilde{f}_{wd}$ for smaller initial deviations, while the initially unaltered stigmator parameters were minimally affected (Supplementary Fig. 3) - both of which are essential conditions for convergence.

Apart from the ability of correcting image aberrations with high accuracy, a well-performing auto-focus algorithm should add minimal computational overhead over the test image acquisition times. We therefore compared DeepFocus processing time to microscope image acquisition time for CPU-only and GPU-based inference, run directly on the microscope control computer. GPU-based inference outperformed CPU-only processing by about an order of magnitude, especially for larger input image patches. Importantly, DeepFocus processing times did not add substantial overhead, even for the little optimized CPU-only mode (Fig. 2f), which will allow widespread deployment of the algorithm to standard microscope computers without hardware modifications.

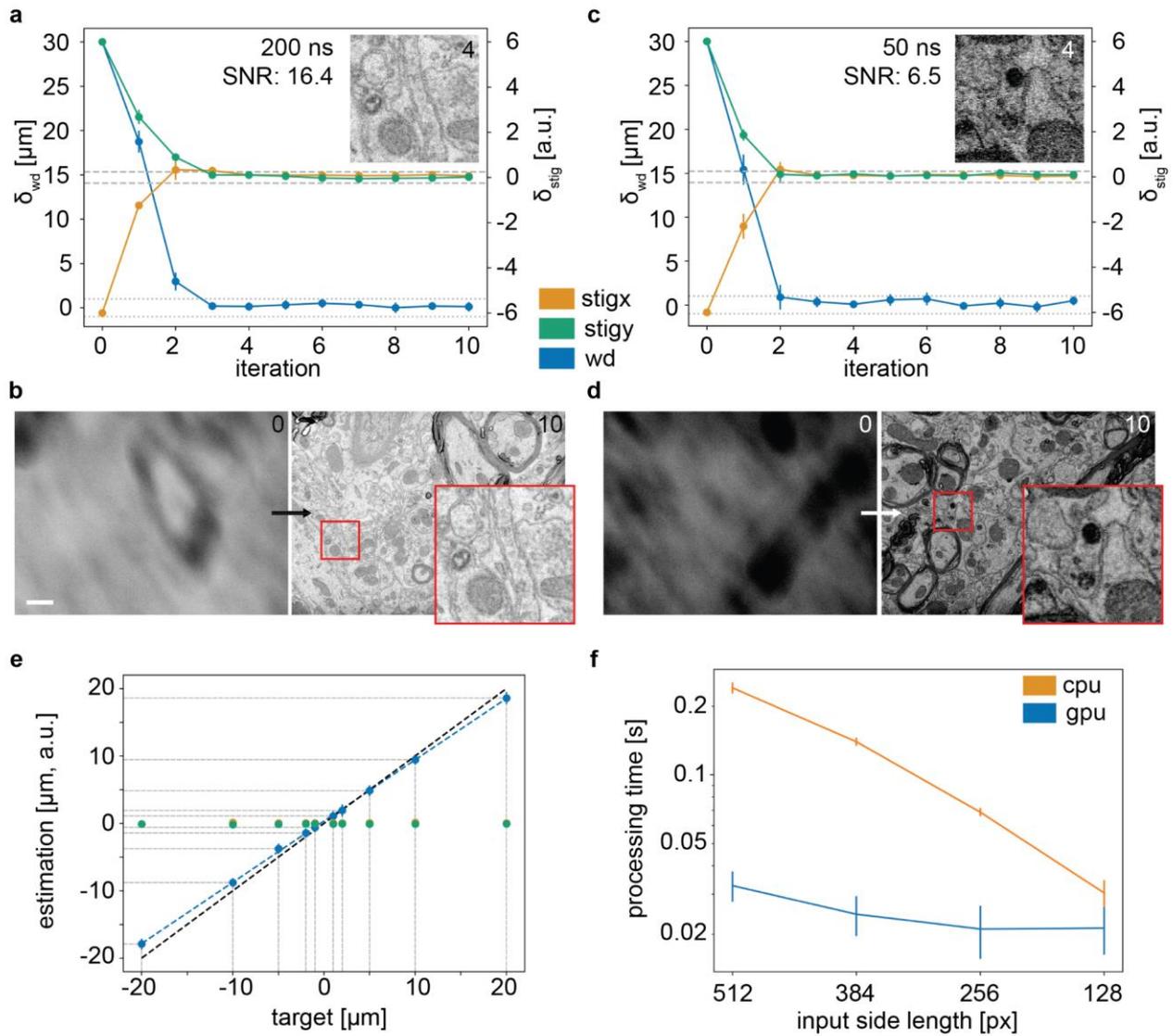

**Fig. 2** DeepFocus convergence and processing time. **a** Convergence plot where each parameter update was calculated as the mean of N predictions with a patch shape of 2 x H x W (height H and width W, in pixels taken from the two perturbed images) using 5 x 2 x 512 x 512 input patches and 200 ns pixel dwell time. The y-axis shows the remaining difference to the initial focus values (dashed and dotted horizontal lines indicate 0.25 and 1 μm margin of stigmator and working distance, cf. Fig. 1b) after each iteration with an initial aberration of 30 μm, +6, -6 (wd, stig x, stig y). The input image size was 1024 x 768 pixels. Numbers in the upper-right corner of the example images indicate the iteration count. The inset plot shows the same region of interest as in b, but taken at 200 ns dwell time. The signal-to-noise ratio (SNR, Methods) was calculated relative to the final focus image after iteration 10 (800 ns dwell time). **b** Image acquired with the initial aberrations and after applying DeepFocus. Scale bar is 1 μm. **c,d** Same as in a,b but with 50 ns pixel dwell time, including the inset in c. **e** Correction estimate (mean and s.d. of 9 different locations; in μm for wd and a.u. for stig x and stig y) after one iteration using 5 × 2 × 512 × 512 input patches with 200 ns pixel dwell time. Colors correspond to those in a and c. **f** Ratio of mean auto focus processing time per input patch-pair to imaging time of the two input images (2 x 769 ms at 2048 × 1536 pixels, 200 ns dwell time) for different patch side lengths (mean and s.d. of 10 repetitions and 10 input patches) on the microscope PC. Error bars represent the uncorrected standard deviation (s.d.).

During DeepFocus development, we noticed that many specimens contain regions with little usable information for an auto-focus algorithm, such as blood vessels in tissue, which show only blank epoxy resin and no contrast that could be used by an auto-focus algorithm (Fig. 3a,b). We therefore reasoned that such areas should have less weight in any $\varDelta F$ estimation, and devised a neural network loss term and architecture that directly leads to the emergence of a second set of model outputs that weigh the $\Delta F$ estimates, without additional training data. These new DeepFocus outputs are loosely regularized (only in terms of weight decay) scores that are used as weighting factors already during DeepFocus model training. We tested two different granularities for weighting, first on the level of the DeepFocus input image patches (Fig. 3c-f), which are cropped portions of the larger input image pair acquired by the microscope, and second, on the level of individual pixels, leading to a scoring of every location in an input image (Fig. 3g,h). Both approaches proved more robust toward specimen regions with little contrast information, demonstrating that DeepFocus does not require potentially error-prone conventional image processing to pre-filter low-contrast regions.

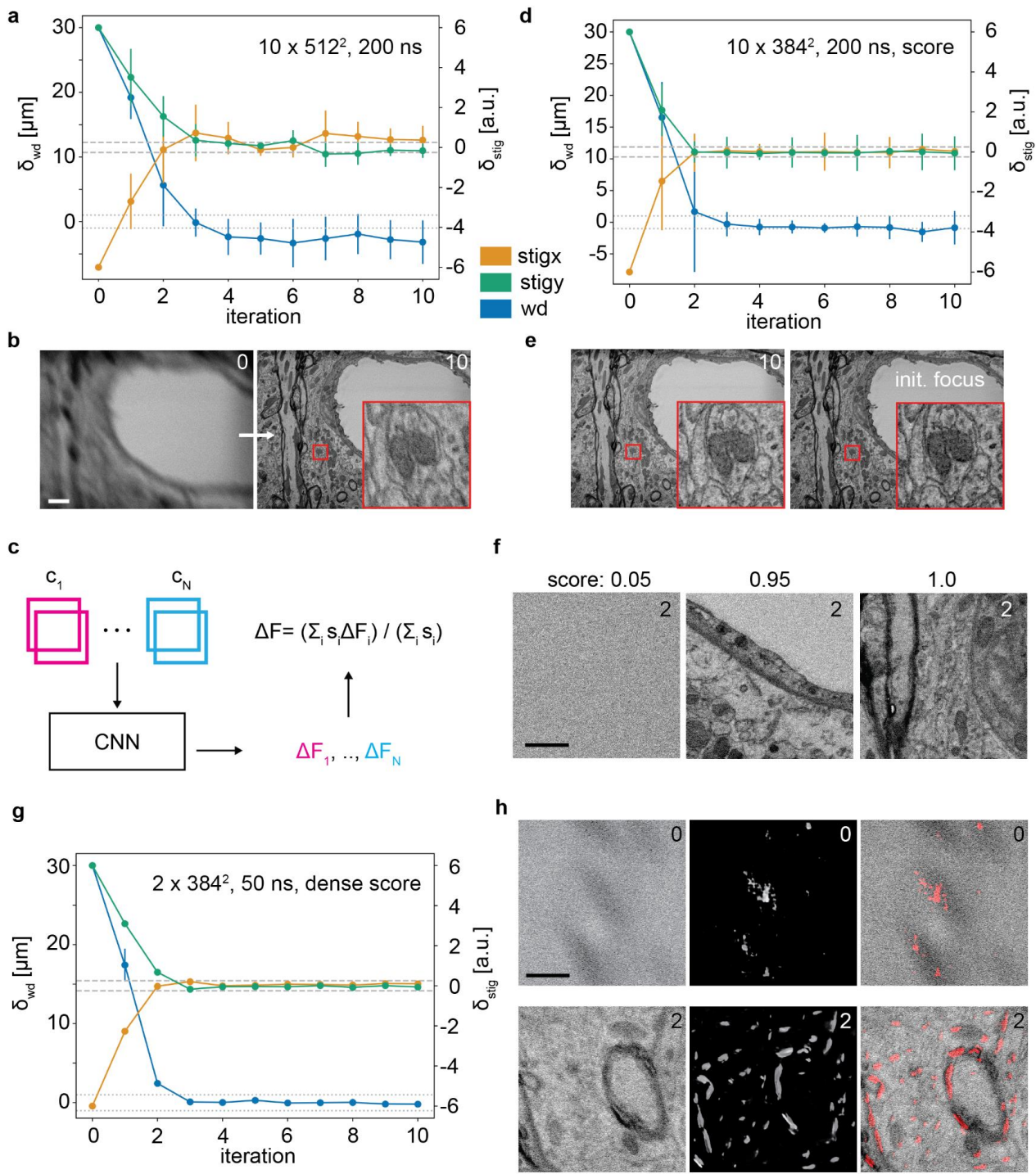

**Fig. 3** DeepFocus model with an additional image region score prediction used to calculate a weighted average estimator for the focus correction. **a** Convergence of the model from Fig. 2a (using 10 instead of 5 patches) on the image in **b** that largely contained a blood vessel with a 200 ns pixel dwell time and an input resolution of 2048 x 1568. **c** Model architecture that predicts an additional score $s_i$ per patch-pair. **d** Convergence of the patch-score model with 10 x 2 x 384 x 384 input patches with the same location and settings as in a. **e** Resulting image using the score model in d and the focused image using the baseline parameters. **f** Sample score values for patches used in iteration 2 (ratios of maximum values; original values: 0.0065, 0.1235, 0.129) together with one of the two input patches. **g** Convergence of DeepFocus with pixel-level score predictions using 2 x 2 x 384 x 384 input crops at 50 ns pixel dwell time and an input image resolution of 2048 x 1536. **h** Score map of one example patch used in g. The right column shows the composite images of the example input patch (left column) and the corresponding pixel scores (center column) in red at iteration 0 and 2. Scale bars: 2 μm in b and 0.5 μm for f and h.

Like MAPFoSt (Maximum-A-Posteriori Focusing and Stigmation)[7], several aberration correction algorithms were developed for SEM in the past, and microscope manufacturer software usually includes such algorithms. In our experience however, these algorithms performed often poorly[7], possibly due to overfitting their parameters, or even the entire algorithmic model to particular test cases. To assess the extent to which DeepFocus is susceptible to overfitting to its remarkably small training set, we first evaluated it on an unseen, non-biological sample and second, on an entirely different microscope, with different imaging settings. Remarkably, DeepFocus generalized exceptionally well to this novel sample (Fig. 4a,b), even with being trained only on image data of a single specimen. Transferring the algorithm to a different microscope with vastly different imaging settings (modified landing energy, beam current, overall working distance range, rotated image acquisition) led to failure and divergence of the model, as expected (Fig. 4c).

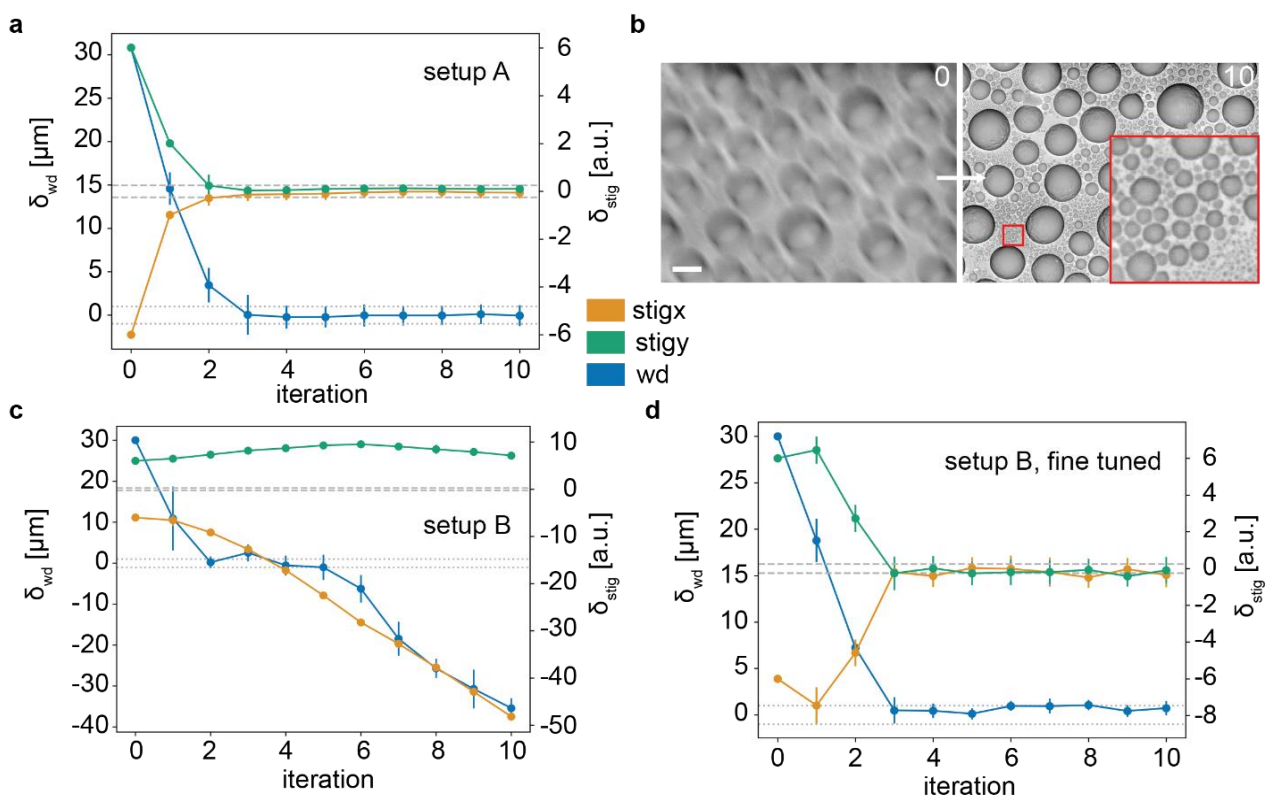

**Fig. 4** DeepFocus convergence on an unseen sample and recalibration for a different setup. **a** Model convergence (same model as in Fig. 3d) on a sample of tin on carbon (not contained in the training data) using setup A at 100 ns dwell time. **b** Image from a at iterations 0 and 10. Scale bar is 1 μm. **c** The same model as applied to tin on carbon on setup B (see Methods for imaging parameters). **d** Convergence of the fine tuned DeepFocus model (last three fully connected layers re-trained) after 50k training iterations using 100 automatically acquired samples at 10 different locations with setup B (Methods).

We therefore developed an alternative approach, DeepScore, aiming for machine and setting independence, by estimating the magnitude of $\Delta F$ without its correction direction from a single image (Methods, see [19,20]). The intent was to create a slower yet machine-independent auto-focus algorithm, based on the directionless score and classical optimization (tested with the simplex method developed by Nelder and Mead[21]). This algorithm can then be used to generate, with minimal manual input, a new training set in case of a required DeepFocus re-calibration. We found that DeepScore, when used with classical optimization, can effectively infer a parameter set $F$ that leads to sharp image formation, albeit, as expected, with slower convergence than the regular DeepFocus model (Supplementary Fig. 4, Supplementary Text 1). Using this approach, we generated a new, smaller training data set (n=10 locations, 31% of the original training set) for a SEM where DeepFocus had diverged. Fine tuning the DeepFocus model (recalibration) took less than 2 hours on a single GPU, and recovered its ability to estimate $\Delta F$ with the original convergence speed (Fig. 4d).

We finally performed a direct comparison of DeepFocus and the state-of-the-art automatic aberration correction algorithm for SEM, MAPFoSt. MAPFoSt uses a Bayesian optimal approach to infer the target $\Delta F$ values, and was specifically optimized to yield a parameter set for sharp images with as little electron dose as possible for the sample. We used the publicly available Python implementation of the algorithm (https://pypi.org/project/mapfost/), with parameters adjusted by its developer (RS) for the SEM used. As expected, MAPFoSt was also able to estimate a correct parameter set on the tested samples (Fig. 5a), but required on average 4 more iterations to convergence ($residual_{wd}$ mean and s.d. of DeepFocus after iteration 2: 0.34 µm ±0.3 µm vs MAPFoSt after iteration 6: 0.5 µm ± 0.21 µm) despite using 50 ns pixel dwell time for the two perturbed images with DeepFocus, and 200 ns for MAPFoSt. Strikingly, DeepFocus outperforms MAPFoSt in particular for low SNR imagery, the image settings domain it was developed for, and large initial aberrations (Supplementary Fig. 5). We also observed that MAPFoSt required longer computation times, more than 30 times, in comparison to DeepFocus running on a low-power GPU inside the microscope computer (processing time per $512^2$ patch-pair with GPU: 0.032 s ± 0.004 s and CPU: 0.240 s ± 0.011 s compared to MAPFoSt with 0.897 s ± 0.024 s for $512^2$ and 1.673 s ± 0.018 s for $768^2$; Methods).

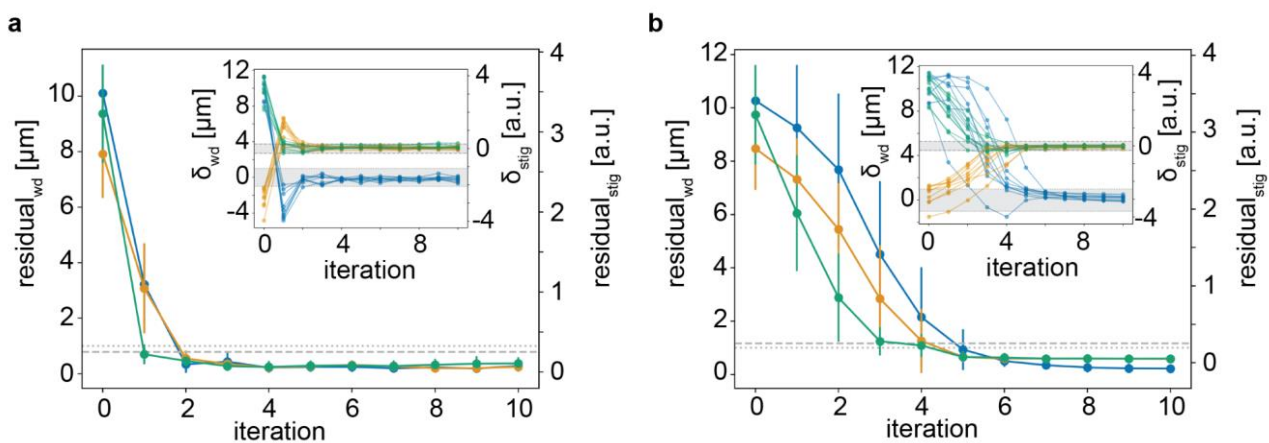

**Fig. 5** Residual error (mean absolute difference from the baseline, see Methods) of nine convergence trajectories of DeepFocus and MAPFoSt using 2048×1536 input images. The individual trajectories are shown in the inset images. **a** DeepFocus model from Fig. 3d with 10 × 2 × 384 × 384 patches and 50 ns pixel dwell time. **b** MAPFoSt with 4 × 2 × 768 × 768 patches and 200 ns pixel dwell time. Colors are consistent with Fig. 2a. Dashed and dotted horizontal lines indicate 0.25 and 1 µm margins for the stigmator and working distance, respectively.

## Discussion

While deep learning has demonstrated impressive advances in recent years in domains such as natural language processing[22] or computer vision[23,24], many "simple" control theory problems remain to be explored[25]. Here we demonstrate how a powerful and overparameterized model, in the classical sense, can outperform the carefully hand-optimized state-of-the-art approach[7] in all measured performance metrics: robustness toward low SNR images, convergence speed, measured by algorithm iterations, and surprisingly, the calculation duration of inference.

This may not be unexpected, given the success of convolutional neural networks across various domains of computer vision, and the fact that auto-focusing can be framed as a regression problem given two input images with known working distance perturbations. We discovered that various neural network architectures, from simple convolutional models followed by fully connected layers to more modern U-Nets[26] were able to solve the problem. This suggests that innovation in machine control may shift from carefully crafting models toward carefully connecting and interfacing more general models.

We believe that the alternative approaches to aberration correction in SEM rely on many implicit and explicit assumptions about the nature of the input images, the electron optics, the point spread function, and, in general, the entire system that is being controlled. While these assumptions are clearly necessary to build a control system based on explicit physical models or classical image processing, they inevitably result in an approximation of the system's behavior. DeepFocus also approximates the system's behavior, but with fewer hard assumptions, and leads to an autofocusing algorithm that is tailored to the peculiarities of every SEM/sample after a simple recalibration, while still generalizing surprisingly well to unseen samples without retraining.

## Methods

### Electron microscopes and samples

All experiments were performed using two different scanning electron microscopes (SEM). The default setup was a Zeiss Merlin SEM equipped with an in-lens secondary electron detector and operated at an acceleration voltage of 1.5 kV, a beam current of 1.5 nA, and a working distance of 4.5 mm (setup A). Recalibration experiments were carried out on a Zeiss UltraPlus SEM with an acceleration voltage of 1.2 kV, a 60 µm aperture, an in-lense secondary electron detector, a working distance of 6 mm, and scans that were rotated by 90° (setup B).

Experiments involving biological samples were conducted on 250 nm sections collected on a silicon wafer. The sections were cut from a 500 µm diameter biopsy punch of a 200 µm thick zebra finch brain slice, stained with Hua protocol [27] and embedded in Spurr's resin. Experiments with non-biological specimens were carried out on tin on carbon from Agar Scientific (S1937) for both setups A and B.

The stigmator values reported by the microscope software (SmartSEM Version 6.06) were used without additional adjustment or calibration.

### Ground truth generation

To generate training and validation samples, a pair of perturbed images (±5 µm working distance) was acquired relative to a known aberration, which was introduced by changing working distance and both stigmators. An expert SEM user manually adjusted the focus baseline at each location, and perturbed image pairs were acquired for 10 introduced aberration vectors (working distance, stig x, stig y). The values of the aberration vectors were drawn uniformly within a given range. Each training sample consisted of two perturbed images as the model input and the corresponding negative aberration vector as the target. The aberrations were sampled at 23 locations from a working distance range of ±20 µm, and astigmatisms of ±0.5. The perturbed images were acquired at a size of 1024 x 768. For 17 locations, images were acquired at a size of 2048 x 1536 and within ±20 µm (wd) and ±5 (stigmators). Finally, the resulting 400 samples were shuffled and divided into training (80%, 320 samples) and validation (20%, 80 samples) sets.

Model architectures and training

All models were developed and trained using PyTorch[28] 1.9.0 and the open source framework elektronn3 (https://github.com/ELEKTRONN/elektronn3) with mini batches, $L_1$ loss, step learning rate scheduler (a factor of 0.99 every 2000 steps) and the AdamW optimizer [29].

The image-to-scalar architecture used seven convolutional layers (valid convolution; 3D kernels to share weights across the two inputs using a z-kernel size of 1) followed by three fully connected layers. The convolutional layers were constructed as follows: convolution, batch normalization, activation (ReLU), max-pooling, and dropout (rate p=0.1).

The architecture for 2 x 512 x 512 inputs was composed of the following layers, using PyTorch pseudocode:

Conv3D(input channels=1, output channels=20, kernel size=(1, 5, 5), pooling size=(1, 2, 2)),
Conv3D(20, 30, (1, 5, 5), (1, 2, 2)),
Conv3D(30, 40, (1, 4, 4), (1, 2, 2)),
Conv3D(40, 50, (1, 4, 4), (1, 2, 2)),
Conv3D(50, 60, (1, 2, 2), (1, 2, 2)),
Conv3D(60, 70, (1, 1, 1), (1, 2, 2)),
Conv3D(70, 70, (1, 1, 1), (1, 1, 1)),
Linear(input channels=6860, output channels=250), activation,
Linear(250, 50), activation,
Linear(50, 3).

For different input shapes, the parameters of the fully connected layers were adjusted as follows:
- 2 × 128 × 128: Linear(140, 100), Linear(100, 50), Linear(50, 3)
- 2 × 256 × 256: Linear(1260, 250), Linear(250, 50), Linear(50, 3)
- 2 × 384 × 384: Linear(3500, 250), Linear(250, 50), Linear(50, 3)

The model output is a correction vector $\Delta \tilde{F}$ for working distance (in µm) and stig x and y (arbitrary units). The $L_1$ loss was calculated without additional weighting as the value range of the different target types (working distance vs. stigmator) appeared sufficiently similar.

In order to obtain an average estimate of multiple corrections with learned weights, the architecture was modified to produce an output of 4 channels (3 for corrections and a weight score associated with each correction: $\Delta F_i, s_i$) instead of 3. The model was trained by computing the weighted average of 5 predictions using the softmax function for normalization of the scores as weights. During each iteration of the training process, 5 patch pairs were generated from the input, and the resulting model output, which was the weighted average, was compared with the target to calculate the loss.

In the image-to-image case, we employed a 3D U-Net architecture[26] with three planar blocks to facilitate weight sharing between the two input images, same convolution, resize convolutions[30] for the upsampling and group normalization[31]. Our model used 32 start filters and two final 2D conv. layers to project the concatenated channels of the two inputs images to 4 channels per pixel: Conv2D(input channels=64, output channels=20, kernel_size=(1, 1)), activation, Conv2D(20, 4, (1, 1)). A softmax function was applied to the 2D score map output which was then used to calculate the weighted average of the per-pixel predictions. Multiple dense predictions were combined by calculating their mean.

In both score models (image-to-scalar and image-to-image) an additional loss term based on the $L_1$ loss of the individual (either patch- or pixel-wise) predictions was added ($\alpha = 0.25$):

$$\tilde{L} = (1-\alpha)L_1^{final} + \alpha L_1^{individual}$$

Model inputs (gray scale images with intensities between 0 and 255) were rescaled to -1 and 1. Patch pairs (one for each of the perturbed images) were cropped randomly (but with the same offset; except for the independent version) and augmented (independently applied with probability p; all values were drawn from a Normal distribution) with additive Gaussian noise (p=0.75, mean=0, sigma=0.2), random gamma adjustment (p=0.75, mean=1.0, gamma s.d.=0.25; pixel intensities internally rescaled between 0 and 1; $I^* = I^\gamma$) and a random brightness and contrast adjustment (contrast mean=1, contrast s.d.= 0.25, brightness mean=0, brightness s.d.=0.25; $I^* = contrast \cdot (I - I_{mean}) + I_{mean} + brightness$).

Trainings were stopped after validation loss convergence at $1 \cdot 10^6$ iterations (no-score models), $0.5 \cdot 10^6$ (patch-score model) and $0.2 \cdot 10^6$ (pixel-score model).

## Multi-trajectory recordings and MAPFoSt comparison

To evaluate the convergence behavior of our models, we monitored the state of the focal parameters during 10 consecutive iterations at a fixed position, using a known initial aberration. Specifically, we plotted the deviation from the focus baseline for three parameters - working distance, stigmator x, and stigmator y - after each iteration (trajectories).

Experiments with DeepFocus and single trajectories used initial aberrations of (30 µm, -6, 6). To determine the parameter baseline, we first coarsely adjusted the focus manually, and then ran the DeepFocus model with patch scores for three iterations, using a dwell time of 200 ns, an image size of 2048 x 1536, and patches sized at 20 x 384 x 384. We subsequently verified the obtained parameter baseline by visually confirming that it led to sharp images.

The signal-to-noise ratio (SNR) of the image presented in Fig. 2 was determined using the methodology proposed by Sage and Unser[32] and a low-noise image, obtained with 800 ns pixel dwell time, as reference. In the experiments conducted with the UltraPlus (setup B) illustrated in Fig. 4, two iterations of the MAPFoSt algorithm (with a 400 ns dwell time and 4 x 786 x 768 patches) were employed to establish the baseline for the unrotated beam scan. Manual focusing, executed by an expert (PS), was utilized for the 90° rotated scan.

Multi-trajectory plots were obtained at 9 distinct locations, evenly distributed on a grid with 80 µm side length. In addition, the mean absolute error (MAE) was computed for each iteration to estimate the average convergence speed and final variance of the model. The initial focus baseline was established through manual focus adjustment, followed by the application of MAPFoSt twice using a 200 ns dwell time, a resolution of 2048 × 1536, and 768 × 768 patches. This baseline was employed to set the initial aberrations. To account for a minor shift in the target focus (working distance) observed during the final iterations, possibly due to the frequent imaging during the trajectory acquisition, two iterations of MAPFoSt or the patch-score model (in the case of Fig. 5a) were performed post-trajectory recording to obtain a more accurate baseline for plotting trajectories and margins in Fig. 5 and Supplementary Fig. 5a. Patch locations for DeepFocus were chosen randomly, yet with a fixed sequence of seeds, i.e. the same N patch offsets (1 offset per patch pair) were used across all trajectories and iterations. Initial aberrations were uniformly sampled within the following ranges: 8 to 12 µm (working distance), -4 to -2 (stig x), and 2 to 4 (stig y), with a fixed random seed to ensure an identical distribution of aberrations for both MAPFoSt and DeepFocus.

The test locations on the specimen for the 9 trajectories were identical for Fig. 5a and Supplementary Fig. 5a. Error bars were calculated using the uncorrected standard deviation in all plots. All experiments involving MAPFoSt were conducted with version 4.2.1 (https://pypi.org/project/mapfost/4.2.1/).

### Compute hardware and timings

Model training was conducted on a Windows computer equipped with two Nvidia Quadro RTX 5000 graphics processing units (GPUs), an Intel Xeon Gold 6240 central processing unit (CPU) @ 2.60GHz (36 threads) and 768 GB RAM. Inference was executed directly on the microscope computers (setup A/B), and the time measurements were carried out on using the Zeiss Merlin microscope computer (Intel Xeon CPU E5-2609 v2 @ 2.50GHz, 4 threads; 16 GB memory; T1000 GPU). The measurements were performed with either the CPU-only or the CUDA (Compute Unified Device Architecture by Nvidia) backend of PyTorch.

The processing time measurement commenced with the perturbed image pair array and concluded with a single correction vector, encompassing cropping, image normalization, CPU-GPU memory transfers, and mean estimation. Initialization of the PyTorch model was excluded from the measurement, as it is only required once during startup. Serialized versions of the model were stored and loaded with TorchScript. The MAPFoSt implementation utilized multithreading on image patches; for instance, for a 2048 × 1536 input image and a patch size of 768 × 768, four parallel processes were spawned. All timing measurements were conducted with 2048 × 1536 images, and the relative time comparison was calculated based on a cycle time of 0.769 s (corresponding to a pixel dwell time of 200 ns) and computed as the mean of 10 repetitions.

### Recalibration procedure

To automatically generate training data for novel setups (DeepFocus recalibration), a separate neural network was developed with the aim of regressing a generalized and microscope-independent image sharpness score (DeepScore). The model designed to produce such a score for a single image was based on an architecture similar to the image-to-scalar DeepFocus variant, consisting of the following layers:

Conv3D(1, 20, (1, 3, 3), (1, 2, 2)),
Conv3D(20, 30, (1, 3, 3), (1, 2, 2)),
Conv3D(30, 40, (1, 3, 3), (1, 2, 2)),
Conv3D(40, 50, (1, 3, 3), (1, 2, 2)),
Conv3D(50, 60, (1, 3, 3), (1, 2, 2)),
Conv3D(60, 70, (1, 3, 3), (1, 2, 2)),
Linear(2520, 250), activation,
Linear(250, 50), activation,
Linear(50, 2).

The model output comprised two scores: one for the working distance $s_{wd}$ and one for the stigmation $s_{stig}$, which may be used for adjustment later on. The loss was calculated using the $L_1$ distance between the absolute ground truth targets (working distance, stigmator x, stigmator y) and the model outputs. The two, absolute stigmator components of the ground truth were summed prior to the loss calculation with the model output score $s_{stig}$. To generate a single score per image, the minima of N patch predictions (with locations selected randomly using a fixed initial seed) were computed independently for each score type (working distance and stigmation) and subsequently summed without additional weights. The resulting single score was used for all experiments.

In order to transform the image sharpness score (objective function) into a microscope-independent autofocus algorithm, we combined it with the downhill simplex method[21]. This approach minimizes the DeepScore through iterative adjustment of the focus parameters. We adopted F. Chollet's Python implementation of the Nelder-Mead algorithm (https://github.com/fchollet/nelder-mead), with the following extension: If there was no improvement within the last 5 iterations (at most every 5 iterations), the current focus parameters were perturbed with noise drawn from a uniform distribution within the ranges (±2 µm, ±0.5, ±0.5).

The automatic adjustment of the focus parameter at each location was achieved using the Nelder-Mead-DeepScore autofocus with 10 × 2 × 512 × 512 patches cropped from an input image with a 200 ns pixel dwell time and a resolution of 2048 × 1536 pixels. The DeepScore network was trained on the ground truth acquired on setup A (see Ground truth generation). To derive a threshold to be used as a stopping criterion for the downhill simplex method, the focus was adjusted manually once before initiating the procedure. The corresponding sharpness score was then evaluated and multiplied by 1.05.

The training image pairs for the DeepFocus recalibration on setup B were acquired on a regular grid with a resolution of 2048 × 1536 pixels and a dwell time randomly chosen as either 200 ns or 100 ns. The first 10 locations' samples were used for training, each sampled with 10 aberrations (uniformly drawn between ±20 µm, ±5, ±5; 100 location-aberration pairs in total; stopping threshold 0.0014). Recalibration was then performed by fine-tuning the parameters of the last three fully connected layers of a pre-trained DeepFocus model. Fine-tuning employed the training parameters described for the DeepFocus, except for an increased learning rate decay, which was achieved by multiplying the rate by 0.95 every 1000 steps and limiting training to a maximum of 50,000 steps (approx. 2 h).

## Data and code availability

All data and source code will be made publicly available on GitHub upon publication.

## Author contributions

PS performed all experiments and conceived the DeepFocus & DeepScore algorithm jointly with JK. RS developed and calibrated the MAPFoSt implementation. The manuscript was written by PS & JK, with contributions by RS.

## Acknowledgements

We would like to thank Winfried Denk for generously providing lab resources from his department. All funding was provided by the Max Planck Society.

## Competing interests

A patent application (EP21212051) covering the method described in this manuscript is currently pending. The authors PS and JK are named as inventors.

# Supplementary Information

## DeepFocus: Fast focus and astigmatism correction for electron microscopy

Schubert et al.

## Supplementary Texts

### Supp. Text 1

We performed additional convergence experiments on setup A where we tested 14 mild test aberrations sampled from a uniform distribution with value ranges between ±10 µm, ±1, ±1 (wd, stig x, stig y; example traces shown in Supp. Fig. 3a), of which all 14 trials converged. The current focus parameters were perturbed with noise drawn from a uniform distribution within (±2 µm, ±0.5, ±0.5) in case there was no improvement within the last 5 iterations (at most every 5 iterations). The focus baseline was found using the model from Fig. 2a with N=10 patches. The stopping threshold was found by scoring the auto-focused image and multiplying it by 1.1 and restricting it to >= 0.001 (all such obtained scores were found to be between 0.001 and 0.0014).

## Supplementary Figures

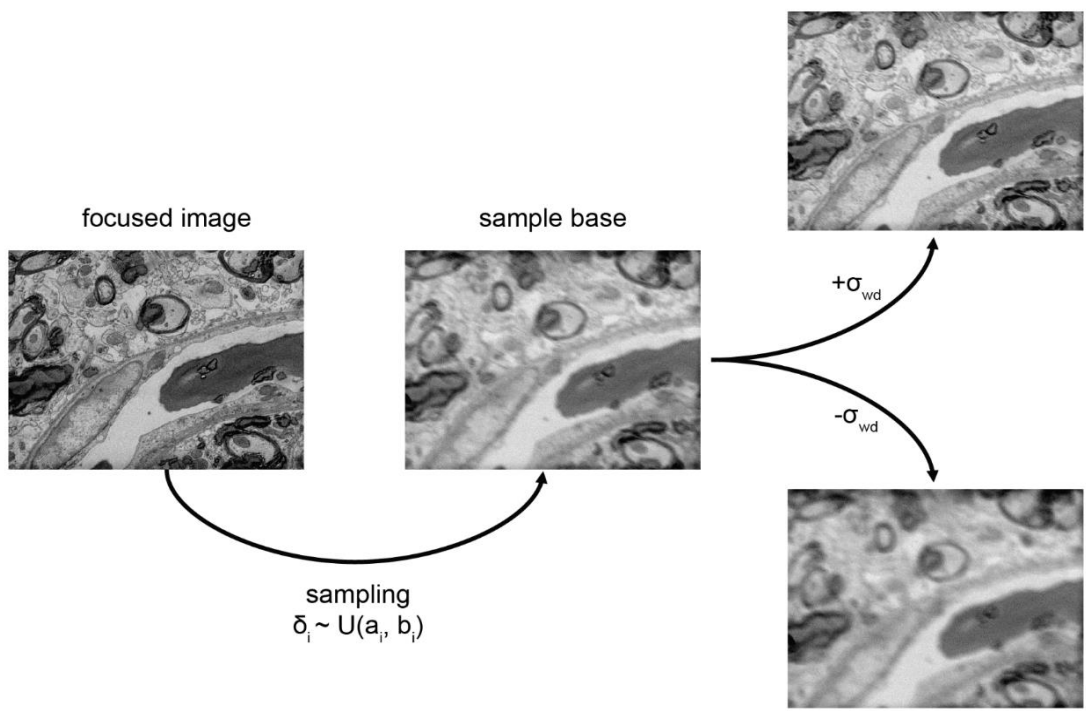

**Supp. Fig. 1** Training and test data set generation. The focus and stigmator values of the focused image are changed by adding a uniformly and independently sampled offset to generate a set of distorted images and corresponding target values.

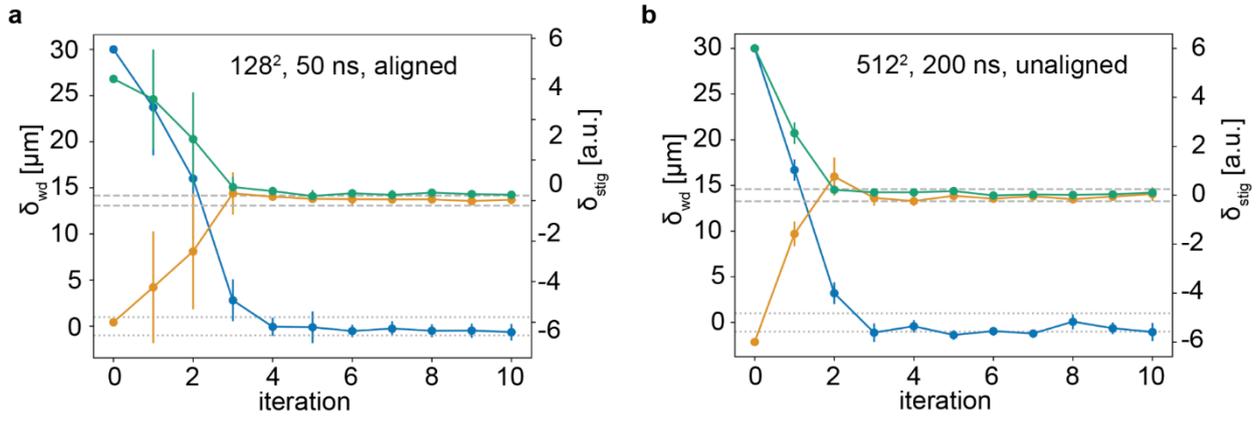

**Supp. Fig. 2** Convergence of DeepFocus using different input properties. **a** 20 x 2 x 128 x 128 input crops and 50 ns pixel dwell time. **b** 5 x 2 x 512 x 512 unaligned input crops, 200 ns pixel dwell time. Crop locations were drawn independently for each perturbed image.

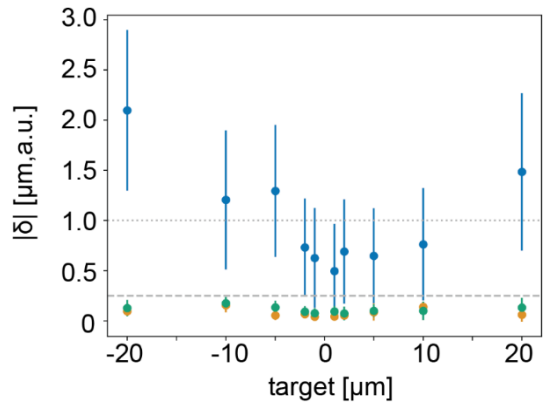

**Supp. Fig. 3** DeepFocus single-iteration performance as a function of initial defocus. Remaining residual error $|\delta|$ between estimate and target from Fig. 2e. Colors as in Fig. 2a. Error bars show the uncorrected standard deviation.

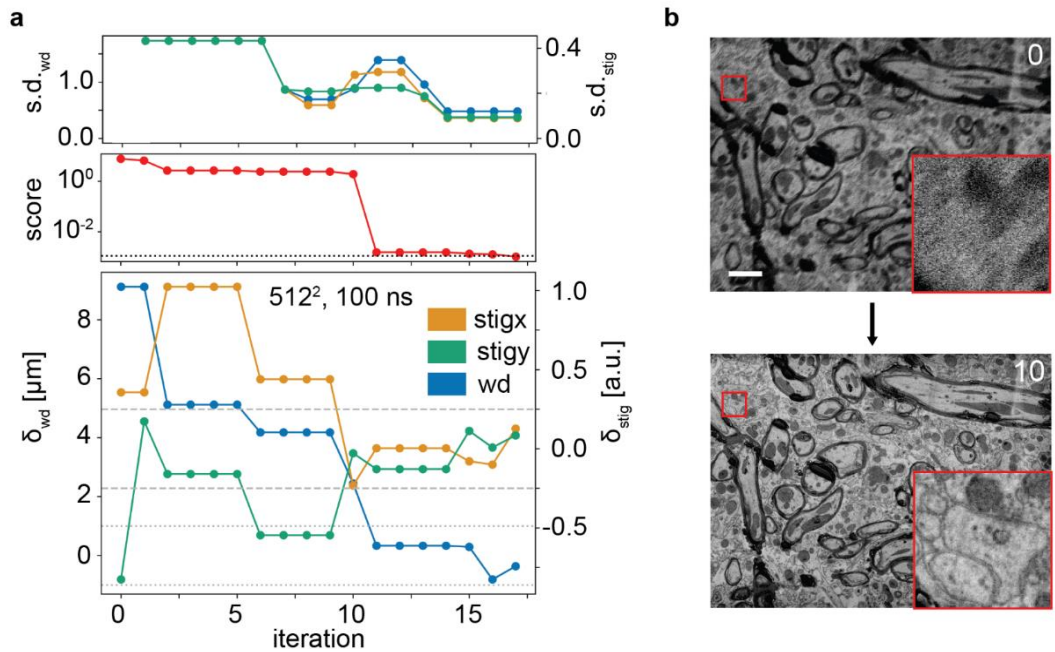

**Supp. Fig. 4** DeepScore auto-focus evaluation. **a** Convergence of the Nelder-Mead optimization using the DeepScore prediction with 5 x 512 x 512 input crops, 100 ns dwell time, input image size of 2048 x 1568 and a total of 37 score evaluations, i.e. image acquisitions. The s.d. (µm for wd and a.u. for stigmators) was calculated from the simplex vertices for each iteration and parameter (iteration 0 is undefined). **b** Sample images used during the Nelder-Mead optimization with the DeepScore objective function from Fig. 4a at iteration 0 and iteration 10 (introduced aberration: 9.11 µm, 0.35, -0.82).

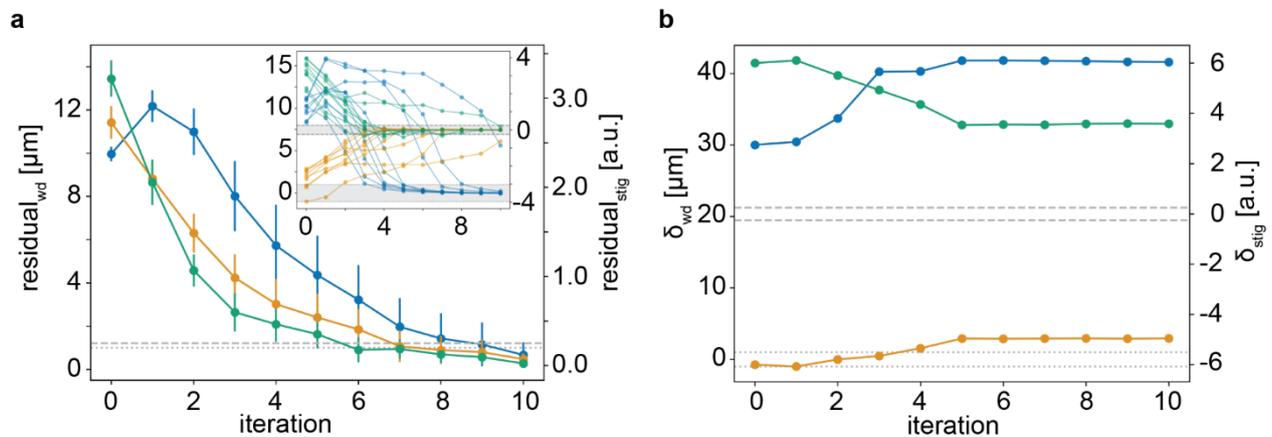

**Supp. Fig. 5** MAPFoSt convergence with challenging initial defocus parameters, using 4 x 2 x 768 x 768 patches and 2048 x 1536 pixel microscope images. **a** Convergence traces with 50 ns pixel dwell time. Initial aberrations were drawn as in Fig. 5. Error bars are ⅓ of the uncorrected standard deviation. **b** Initial aberration (30 µm, -6, 6) and 200 ns dwell time.